\documentstyle[12pt]{article}
\begin{document}
\newcommand{\be}{\begin{equation}}
\newcommand{\ee}{\end{equation}}
\newcommand{\bea}{\begin{eqnarray}}
\newcommand{\eea}{\end{eqnarray}}
\baselineskip=18pt
\begin{center}
{\large{\bf 
Time-Dependent Variational Principle for $\phi^4$ Field Theory: \\
RPA Approximation and Renormalization (II)$^{\ast}$ }}
\end{center}
\vspace{0.5cm}
\baselineskip=15pt
\begin{center}
Arthur K. Kerman 
\end{center}
\begin{center}
{\it Center for Theoretical Physics \\
     Laboratory for Nuclear Science \\
     and Department of Physics \\
     Massachussets Institute of Technology \\
     Cambridge, Massachusetts\ \ 02139\ \ \ U.S.A.} 
\end{center}
\begin{center}
and
\end{center}
\begin{center}
Chi-Yong Lin$^{\dag}$
\end{center}
\begin{center}
{\it Instituto de F\'{\i}sica, Universidade de S\~ao Paulo \\
     Caixa Postal 66318, 05315-970, S\~ao Paulo\\
     S\~ao Paulo \ \ SP \ \ \ Brazil} 
\end{center}

\baselineskip=15pt
\vspace{0.5cm}
\begin{center}
{\bf ABSTRACT}
\end{center}

\indent The Gaussian-time-dependent variational equations are used to
explored the physics of $(\phi^4)_{3+1}$ field theory. We have
investigated the static solutions and discussed the conditions of
renormalization. Using these results and stability analysis we show that
there are two viable non-trivial versions of $(\phi^4)_{3+1}$. In the
continuum limit the bare coupling constant can assume 
$b\rightarrow 0^{+}$ and $b\rightarrow 0^{-}$,
which  yield well defined asymmetric 
and symmetric solutions respectively. We have also
considered small oscillations in the broken phase and shown that they give
one and two meson modes of the theory. The resulting equation has a
closed solution leading to a ``zero mode'' and vanished scattering
amplitude in the limit of infinite cutoff.

\bigskip
\centerline{\it To be published in Annals of Physics}
\vfill

\hspace{\fill}
March, 1998

\noindent\makebox[66mm]{\hrulefill}

\footnotesize 
$^{\ast}$This work is supported in part by funds
        provided by the U. S. Department of Energy (D.O.E.) 
        under cooperative agreement\#DE-FC02-94ER40818.


$^{\dag}$
Supported by Fun\c{c}\~ao de Amparo \`a Pesquisa do Estado de S\~ao
Paulo (FAPESP), Brazil.

\normalsize
\newpage
\baselineskip=30pt
\renewcommand{\theequation}{1.\arabic{equation}}
\setcounter{equation}{0}
\begin{center}
{\bf I. Introduction}
\end{center}

\indent
In a recent paper [1](hereafter referred to as I)
we have obtained the RPA equations for
$\phi^4$ field theory by linearizing the time-dependent variational
equations. The method was implemented   
for the case of symmetric vacuum, $\langle\phi\rangle=0$, which allows
us to investigate two-meson physics. 
We have shown that it is a simple 
nonperturbative method to study
scattering processes. Using this framework the problem of 
stability of vacuum can be
explored from the RPA modes.  
In continuation of
I we will consider here the  stability of the theory for other critical
points of the Gaussian parameter space[2]. In particular, we discuss
the solutions and renormalization conditions for the asymmetric
vacuum.
In this case, the excitations of the vacuum are identified with 
the one and two-particle wavefunctions
and the system of equations can also be solved analytically. 
As result, a stable zero mode is found
for certain range of
renormalized coupling constant and as well as a complete form of the
scattering amplitude.

\indent
For completeness, let us first repeate here the key equations of 
I. 
The bare parameter hamiltonian for the $\phi^4$ theory is          
[We use the notation: $\int_{\bf x} = \int d^3 x$]
\be
        \hat H=\int_{\bf x}\Bigl({1\over 2}\hat\pi^2({\bf x})
                             +{1\over 2}(\nabla\hat\phi({\bf x}))^2
                             +{a\over 2}\hat\phi^2({\bf x})
                             +{b\over 24}\hat\phi^4({\bf x})\Bigr) \, .
\ee
As an approximation, we take a Gaussian trial wave functional
\bea
\psi(\phi,t)=N exp \Biggl\{&-&\int_{\bf x,y}
                   \Bigl[\phi({\bf x})-\phi_0({\bf x},t)\Bigr]
                     \Bigl[{G^{-1}({\bf x,y},t) \over 4}
                        -i\Sigma({\bf x,y},t)\Bigr]
                     \Bigl[\phi({\bf y})-\phi_0({\bf y},t)\Bigr] \nonumber\\
                    &+&i\int_{\bf x} \pi_0({\bf x},t)
                        \Bigl[\phi({\bf x})-\phi_0({\bf
                          x},t)\Bigr]\Biggr\}\, ,
\eea
where $N$ is a normalization and $\phi_0({\bf x},t), \pi_0({\bf x},t),
G({\bf x,y},t)$ and $\Sigma({\bf x,y},t)$ are our variational parameters.
These quantities are related to the following 
mean-values:
\be
        \langle\psi,t\mid\hat\phi({\bf x})\mid\psi,t\rangle 
        =\phi_0({\bf x},t) \, ,
\ee
\be
        \langle\psi,t\mid\hat\pi({\bf x})\mid\psi,t\rangle  
        =\pi_0({\bf x},t) \, ,
\ee
\be
        \langle\psi,t\mid\hat\phi({\bf x})\hat\phi({\bf y})\mid\psi,t\rangle 
        =G({\bf x,y},t)+\phi_0({\bf x},t)\phi_0({\bf y},t) \, ,
\ee
\be
        \langle\psi,t\mid i{\delta \over \delta t}\mid\psi,t\rangle
        =\int_{\bf x,y}\Sigma({\bf x,y},t)\dot G({\bf x,y},t)
          +\int_{\bf x} \pi_0({\bf x},t)\dot\phi_0({\bf x},t) \, .
\ee
From (1.1) and (1.2) one can compute the effective hamiltonian,
\be
        {\cal H}=\langle\psi,t\mid\hat H \mid\psi,t\rangle ,
\ee
which is the energy of the system,
\bea
{\cal H}&=&\int_{\bf x}\Bigl[\frac{1}{2}\pi^2_0({\bf x},t)
                           +\frac{1}{2}(\nabla\phi_0({\bf x},t))^2
                           +\frac{a}{2}\phi^2_0({\bf x},t)
                           +\frac{b}{24}\phi^4_0({\bf x},t)\Bigl]\nonumber\\
        &+&\frac{1}{8}\int_{\bf x} G^{-1}({\bf x,x},t)
      +2\int_{\bf x,y,z}\Sigma({\bf z,x},t)G({\bf x,y},t)\Sigma({\bf y,z},t)
                                                                 \nonumber\\
        &+&\frac{1}{2}\int_{\bf x}[-\nabla^2_{\bf x}
                                   +a+{b \over 2}\phi^2_0({\bf x},t)]
                                   G({\bf x,y},t)|_{\bf x=y} 
        +\frac{b}{8}\int_{\bf x} G({\bf x,x},t)G({\bf x,x},t) \, .
\eea
The variational equations of motion read
\bea
\dot\phi_0({\bf x},t) &=& \pi_0({\bf x},t) \, ,\\
\dot\pi_0({\bf x},t) &=& \nabla^2\phi_0({\bf x},t)
                        -a\phi_0({\bf x},t)
                        -\frac{b}{6}\phi_0^3({\bf x},t)
                   -\frac{b}{2}\phi_0({\bf x},t)G({\bf x,x},t) \, ,\\
\dot G({\bf x,y},t) &=& 2\int_{\bf z}\Bigl[G({\bf x,z},t)\Sigma({\bf z,y},t)
                        +\Sigma({\bf x,z},t)G({\bf z,y},t)\Bigr] \, ,\\
\dot\Sigma({\bf x,y},t) &=& -2\int_{\bf z}\Sigma({\bf y,z},t)\Sigma({\bf z,x},t)
        +\frac{1}{8}G^{-2}({\bf x,y},t) \nonumber\\
        &+&\frac{1}{2}\Bigl[\nabla^2_{\bf x}-a-{b \over 2}\phi^2_0({\bf x},t)
        -\frac{b}{2}G({\bf x,x},t)\Bigr]\delta({\bf x-y}) \, .
\eea

\indent
These are nonlinear time-dependent field equations. Therefore,
a closed solution is not easily constructed. Here, we will 
consider the equations in the equilibrium situation and the small
oscillation regime. In these cases, an explicit solution can be
obtained allowing us to examine diverse properties of the theory.
The structure of this paper is as follows. Section II discusses the 
time independent Gaussian equations. Particular attention is
paid to the question of the renormalization and we study 
solutions for the resulting gap equation. We also discuss 
the so called Gaussian effective potential for the case when the bare
coupling constant
$b\rightarrow 0^{+}$. Section III 
discusses stability of the stationary points found in the
section II by investigating the properties of variational space. In Section
IV we shall derive the RPA equation by considering near equilibrium
dynamics about the critical points. Since the
excitations of vacuum are quantum particles, Section V will solve the
RPA equation with boundary conditions of scattering processes. In Sec VI
we will use the properties of separable potential 
to get solutions for the $\bf T$
matrix. The spectrum of RPA modes is discussed using this result and
conditions of stability will be analysed within this context.

\renewcommand{\theequation}{2.\arabic{equation}}
\setcounter{equation}{0}
\begin{center}
{\bf II. Time-Independent Gaussian Variational Equations}
\end{center}

\indent 
This section will discuss the variational equations, Eqs.(1.8)-(1.12), in the 
equilibrium situation. We investigate the solutions of these equations
and study the renormalization conditions. We will
consider $\phi_0$ uniform, $\phi_0({\bf x})\equiv\varphi$,  
and  parametrize the kernel $G$ as
${1\over4}G^{-2}({\bf x},{\bf y})
\equiv\langle {\bf x}\mid \hat p^2 + m^2\mid{\bf y}\rangle$.
Therefore, the parameters $\varphi$ and $m$ are taken to be 
independent of ${\bf x}$.  
With these assumptions the expectation value of $H$,
(1.8), with kinetic term set to zero, reads as
\bea
\label{II1}
U(\varphi,m)&=&\int_{\bf x} u(\varphi,m) \, ,\nonumber \\
u(\varphi,m)&=&{1\over4}G^{-1}(m)-{m^2\over2}G(m) 
              +{1\over 2b}\Bigl[a + \frac{b}{2} G(m)\Bigr]^2 \nonumber \\
     & &+{1\over2}\Bigl[a + \frac{b}{2} G(m)\Bigr]\varphi^2
     +{b\over24}\varphi^4 \, .
\eea
In this equation $G(m)$ and $G^{-1}(m)$ are defined as
$\langle {\bf x}\mid G \mid {\bf x} \rangle$ and 
$\langle {\bf x}\mid G^{-1} \mid {\bf x} \rangle$ respectively.
By virtue of parametrization of $G$ these matrix elements are given by
\be
\label{II2}
G(m)=\langle {\bf x}\mid {1\over 2\sqrt{\hat p^2+m^2}}
                    \mid {\bf x}\rangle 
    ={1\over 8\pi^2}\left[\Lambda_{\bf p}^2
      -m^2\log\bigl({2\Lambda_{\bf p}\over\sqrt{e}m}\bigr)\right] \, ,
\ee
\be
\label{II3}
G^{-1}(m)=\langle {\bf x}\mid 2\sqrt{\hat p^2+m^2}
                      \mid {\bf x}\rangle 
={1\over
  8\pi^2}\left[2\Lambda_{\bf p}^4+2m^2\Lambda_{\bf p}^2
  -m^2\log\bigl({2\Lambda_{\bf p}\over\sqrt{e}m}\bigr)-{m^4\over
                      4}\right] \, ,
\ee
where $\Lambda_{\bf p}$ is a cutoff in the integrals.
Notice that the energy density $u$ is a function of two variables
$\varphi$ and $m$. The next step  is to obtain the critical points of
this functions and 
find the renormalization conditions. 

\smallskip
\begin{center}
{\bf II--a. Gap Equation and Renormalization}
\end{center}
\smallskip

Minimization of $u$ with respect to $\varphi$ and $m$ yields
\be
\label{II4}
{\partial u\over\partial\varphi}=
\varphi\Bigl( a +\frac{b}{6}\varphi^2
+\frac{b}{2}G(m)\Bigr)=0 \, ,
\ee
\be
\label{II5}
{\partial u\over\partial m}=
mG'(m)
\Bigl[a+{b \over 2}\varphi^2
+\frac{b}{2}G(m)-m^2\Bigr]=0 \, ,
\ee
\noindent
where $G'(m)=dG(m)/dm^2$. Notice that one may have several sets of
solutions for the system of equations (2.4-2.5).

\indent By analysing the 
stability matrix (see section III) one can show that 
the solution $m=0$ of (2.5) is always a saddle point of $u(\varphi,m)$ 
because the determinant is negative in this case. 
Thus, the possible stable points of (2.1) are determined
through the gap equation, defined by
\be
\label{II6}
m^2=a+{b \over 2}\varphi^2
+\frac{b}{2}G(m) \, ,
\ee
and (2.4). From its solutions, $\varphi=0$ and 
\be
\label{II7}
a +\frac{b}{6}\varphi^2+\frac{b}{2}G(m)=0,
\ee
one can eliminate $\varphi$
in favor of $m$ and arrive at the following key equations to be solved,
\be
\label{II8}
m^2=\alpha\Bigl( a+\frac{b}{2} G(m)
          \Bigr) \, .
\ee
In this expression, 
we have parametrized the two solutions of (2.4) as follows: $\alpha=1$
when $\varphi=0$ and $\alpha=-2$  for $b\varphi^2=-6[a+b/2G(m)]$. 
On the other hand, by combining (\ref{II6}) and (\ref{II7}) 
it is easy to show that
\be
\label{II9}
b\varphi^2=3m^2
\ee
which we define as the broken phase. Notice that this phase might 
degenerate with the massless symmetric 
phase if $m=0$ is a solution of (\ref{II8}). We will come back to 
this point later.

%
%
\indent Let us now examinate the
renormalization conditions for the gap equation (\ref{II8}). First, we
use (\ref{II2}) to rewrite (\ref{II8}) as
\be
\label{II10}
{m^2\over\alpha}=a+{b\Lambda_{\bf p}^2\over18\pi^2}
          -{bm^2\over16\pi^2}\log\frac{2\Lambda_{\bf p}}{\sqrt{e}m}\, .
\ee
Next we introduce an arbitrary  mass scale $\mu$ by writing
$\log(\Lambda_{\bf p}/m)=\log(\Lambda_{\bf p}/\mu)+\log(\mu/m)$, the equation
(\ref{II10}) becomes
\be
\label{II11}
m^2=\frac{{\displaystyle a+\frac{b\Lambda_{\bf p}^2}{16\pi^2}}}
         {{\displaystyle \frac{1}{\alpha}
          +\frac{b}{16\pi^2}\log\frac{2\Lambda_{\bf p}}{\sqrt{e}\mu}} }
   +\frac{{\displaystyle \frac{b}{32\pi^2}}}
         {{\displaystyle \frac{1}{\alpha}
          +\frac{b}{16\pi^2}\log\frac{2\Lambda_{\bf p}}{\sqrt{e}\mu}} }
           m^2\log\frac{m^2}{\mu^2}\,.
\ee
This equation renders finite result if we choose the following 
self-consistency condition 
\bea
\label{II12}
\epsilon\mu^2&=&\frac{{\displaystyle a+\frac{b\Lambda_{\bf p}^2}{16\pi^2}}}
         {{\displaystyle \frac{1}{\alpha}
         +\frac{b}{16\pi^2}\log\frac{2\Lambda_{\bf p}}{\sqrt{e}\mu}}},
         \hspace{1.0cm} \epsilon=0,\pm1, \\
\vspace{0.3cm}
\label{II13}
{1\over g_{\mu}}&=&\frac{2}{\alpha b}+L_{\mu}, \hspace{1.0cm}
          L_{\mu}=\frac{1}{8\pi^2} \log\frac{2\Lambda_{\bf p}}{e\mu} .
\eea
Thus, the renormalized version of Eq.(\ref{II8}) reads
\be
\label{II14}
m^2=\epsilon\mu^2+\frac{ \bar g_{\mu}}
{1+\bar g_{\mu}} m^2\log\frac{m^2}{\mu^2}\, ,
\ee
where ${\bar g_{\mu}}=g_{\mu}/16\pi^2$ and $\epsilon$ indicates different
renormalization conditions. The above result can be interpreted as
follows. We start with two bare parameters $a$ and
$b$. With the help of the transformations rules 
(\ref{II12})-(\ref{II13}) we arrive at a  
new set of paramteters $\mu$ and $g$. Being
$\mu$ a renormalized mass, which also defines a mass scale, and $g$ a
dimensionless renormalized coupling constant, in a such way that the
resulting gap equation involves finite quantities only.
Notice also from (\ref{II12})-(\ref{II14}) that two different theories of
$(\phi^4)_{3+1}$ are involved. One is the version with $b\propto
1/L_{\mu}$ ($\alpha=1$) and the other is $b\propto
-1/L_{\mu}$ ($\alpha=-2$) which belong to 
two distinct field theories [3-4]. In the limit of
infinite cutoff the renormalized coupling constant 
$g_{\mu}$ can assume any value.

\smallskip
\begin{center}
{\bf II--b. Massive and Massless Solution}
\end{center}
\smallskip

\indent In order to discuss the possible solutions of (2.8) and
investigate the
roles played by $\epsilon$ we compute $u(\varphi(m),m)$, being
$\varphi(m)$ the solutions of (2.4), 
\bea
\label{II15}
u(m)&=&u(\varphi(m),m) \nonumber \\
&=&{1\over4}G^{-1}(m)-{m^2\over2}G(m) 
      -{\alpha\over 2b}\Bigl[a + \frac{b}{2} G(m)\Bigr]^2 \\
%
\label{II16}
&=&\frac{1}{128\pi^2}\Bigl( m^4\log \frac{m^4}{\mu^4}-m^4+\mu^4 \Bigr)
     -\Bigl( \frac{1}{\bar g_{\mu}} + 1 \Bigr)
      \frac{\bigl( m^2-\epsilon\mu^2\bigr)^2 }{64\pi^2}
      \, ,
\eea
where we have used (2.3) and 
the counterterms introduced in (2.12)-(2.13)[1,2].
The different behavior of $u(m)$ with $\bar{g}_{\mu}<0$ and $\epsilon=1$  are
shown in the fig.1a of I. In this case $u(m=\mu)$ and $u(m=0)$ are
local or true minima depeding on the values of $\bar{g}_{\mu}$. However
$m=0$ shown there actually corresponds to the massless solution of 
(\ref{II5}), which is
a saddle point in the $(\varphi,m)$ space. In addition, $u(m)$ calculated
with $\bar{g}_{\mu}>0$ reduce to those  in fig.1a if
one uses the scale $\bar m$ defined as minimum of $u(m^2)$
(see appendix B of I) . 
A similar consideration can be made also for the case
$\epsilon=-1$ as follows.

\indent Recalling (\ref{II13}) we can write a relation of renormalized
coupling constants at mass scales $\mu$ and $\bar{m}$ readly as  
\be
\label{II17}
        {1\over\bar g_{m}}
        = {1\over \bar g_{\mu}}
         -\log {\bar m^2\over\mu^2} \, ,
\ee
where $\bar{m}$ is defined as the minimum of $u(m)$. Combining now
(\ref{II14}) with (\ref{II17}) one
can get the following simple relation
\be
\label{II18}
\bar{m}^2\Bigl(1+{1\over\bar g_{m}}\Bigr)
=-\mu^2\Bigl(1+{1\over \bar g_{\mu}}\Bigr) \, .
\ee
Using this result we can rewrite (\ref{II16}), at  scale of 
$\bar{m}$, as
\be
\label{II19}
{64\pi^2\over\bar{m}^4} u(m)= 
        {m^4\over\bar m^4} \bigl( \log {m^2\over\bar m^2} -{1\over 2} \bigr)
        -\bigl( {1\over\bar g_{m}}+1\bigr)
                         \bigl( {m^2\over\bar m^2}-1\bigr)^2.
\ee
Disregarding some unimportant additive constants this equation is equal
to (\ref{II16}) in unit of $\bar{m}$. Therefore, this renormalization
scheme does not add any new physics.

\indent The case when $\epsilon=0$ corresponds to the renormalization
prescription $a+b\Lambda^2/16\pi^2=0$ [2,5]. 
Thus, the gap equation (2.8) becomes
\be
\label{II20}
m^2\Bigl({1\over\alpha}+{b\over16\pi^2}\log\frac{2\Lambda}{\sqrt{e}m}\Bigr)=0
\ee
which allows a massless solution. It is important to notice that there
is only one free parameter involved in this case. Thus, we may fix it using the
mass scale $\bar{m}$ defined by the second solution of (\ref{II20}), namely
\be
\label{II21}
{2\over\alpha b}
=-{1\over8\pi^2}\log\frac{2\Lambda}{\sqrt{e}\bar{m}}\, .
\ee
Comparing this to (\ref{II13}) and using (\ref{II17}) one has
$\bar g_{m}=-1$. With these ingredients 
we find the following the renormalized version of $u$ for this case:
\be
\label{II22}
u(m)= {m^4\over 128\pi^2} \bigl( \log {m^4\over\bar m^4}-1 \bigr).
\ee
On the other hand, one can obtain the above result rewriting directly
(\ref{II16}) in unit of $\bar{m}$ defined by (\ref{II20}). 
In other words, the case of
$\epsilon=0$ corresponds to the result given by (\ref{II19})
with $\bar{g}_{m}=-1$. 
Therefore, we can conclude after this
analysis that all
possible physics situations in $u(\varphi(m),m)$ are contained in the case of
$g_{m}<0$ and $\epsilon=1$ when $\bar{m}$ is utilized as the mass
scale. It has minima at $m=\bar{m}$, and
massless solution $m=0$, for the particular case
of $\bar g_{m}=-1$. Thus, the time-independent variational equations 
(\ref{II4})-(\ref{II5}) allow simple analytic solution and energy
density $u(\varphi,m)$ has equilibrium points at $(\varphi=0,\bar m)$
and $(b\varphi^2=3\bar m^2,\bar m)$. 

\begin{center}
{\bf II--c. Gaussian Effective Potential}
\end{center}

\indent In the previous subsections we have studied the energy density
$u(\varphi,m)$ when it is minimum with respect to 
$\varphi$, i.e., $u(\varphi(m),m)$, being $\varphi(m)$ given by (2.4). 
Now we want to
calculate $u(\varphi,m)$ when it is minimum with respect to $m$.
In particular, when the resulting expression is written as a function of
$\varphi$, it is known as Gaussian Effective Potential (GEP) [3-4]. 
However, an explicit expression of $m(\varphi)$ from 
(\ref{II6}) is not
straighforward. Instead, we write $\varphi$ in terms of $m$,
\label{II23}
\be
\frac{b\varphi^2}{2}=m^2-a-\frac{b}{2}G(m) \, .
\ee
Substituting this into (\ref{II1}) yields
\be
\label{II24}
u(m)={1\over4}G^{-1}(m)-{m^2\over2}G(m) 
    -\frac{1}{b}\Bigl[a + \frac{b}{2} G(m)\Bigr]^2+D(m) \, ,
\ee
where 
\be
\label{II25}
D(m)=\frac{m^4}{6b}+\frac{2m^2}{3b}\Bigl[ a+\frac{b}{2}G(m) \Bigr]
                +\frac{2}{3b}\Bigl[a+\frac{b}{2}G(m)\Bigr]^2. 
\ee
Notice that $u(m)-D(m)$ is exactly equal to (\ref{II15}) for the of case
$\alpha=-2$. Using now the renormalization procedure given by
(\ref{II12})-(\ref{II13})  it is straightforward to show that 
$
D\rightarrow {\cal O}\Bigl(\frac{1}{L_{\mu}}\Bigr).
$
Therefore, the energy density at
the curves defined by (\ref{II6}) as well as by (\ref{II7}) in the
$(\varphi,m)$ plane has the same result. The coincidence is not casual and let
us look at (2.1) more carefully. Recalling the relation (\ref{II9}) we
note that since $m^2$ is a finite physical
quantity, therefore it suggests that $\varphi$ requires 
a scaling factor. Thus, we define a finite mean-field value as 
\be
\label{II26}
b\varphi^2=\Phi^2 \, .
\ee
Using now (\ref{II12})-(\ref{II13}) one can rewrite (\ref{II6})
and (\ref{II7}) respectively as
\be
\label{II27}
\Phi^2=3m^2+\frac{1}{L_{\mu}}
\Bigl[\frac{m^2-\mu^2}{16\pi^2}\Bigl(\frac{1}{\bar g_{\mu}} +1 \Bigr)
               -\frac{m^2}{8\pi^2}\log\frac{m}{\mu}
            \Bigr]\, ,
\ee
\be
\label{II28}
\Phi^2=3m^2+\frac{3}{L_{\mu}}
\Bigl[\frac{m^2-\mu^2}{16\pi^2}\Bigl(\frac{1}{\bar g_{\mu}} +1 \Bigr)
               -\frac{m^2}{8\pi^2}\log\frac{m}{\mu}
            \Bigr]\, .
\ee
This result shows that in the limit of infinite cutoff, the two curves
converge to $b\varphi^2=3m^2$. Using this relation one can write GEP 
from (\ref{II16}) immediately as
\be
V_{_{GEP}}(\Phi)
=\frac{\Phi^4}{576\pi^2}\Bigl( \log \frac{\Phi^2}{3\mu^2}-\frac{1}{2} \Bigr)
     -\Bigl( \frac{1}{\bar g_{\mu}} + 1 \Bigr)
      \frac{\Phi^2}{192\pi^2}
      \Bigl( \frac{\Phi^2}{3}-2\mu^2\Bigr) \, .
\ee
In particular, (2.29) recovers the  GEP
obtained by the authors of Ref.[3] after some changes of variables 
(cf. their Eq.(17)). 
%
\renewcommand{\theequation}{3.\arabic{equation}}
\setcounter{equation}{0}
\begin{center}
{\bf III. Stability Analysis }
\end{center}

\indent
In this section we shall analyse the stability conditions for the
solutions obtained in the previous discussion. Since the energy
density $u(\varphi,m)$, (\ref{II1}), is a function of two variables, we
can define its stability matrix as
\label{III1}
\bea
A&=&
\left(
\begin{array}{cc}
{\displaystyle\partial^2 u\over\displaystyle\partial\varphi\partial\varphi } &
{\displaystyle\partial^2 u\over\displaystyle\partial\varphi\partial m }  \\
{\displaystyle\partial^2 u\over\displaystyle\partial m \partial\varphi } &
{\displaystyle\partial^2 u\over\displaystyle\partial m \partial m } 
\end{array}
\right) \nonumber \\
&=&
\left(
\begin{array}{cc}
a+{\displaystyle b\over \displaystyle 2}G(m)
 +{\displaystyle b\over \displaystyle 2}\varphi^2 &
b\varphi m G'(m) \\
b\varphi m G'(m) &
\bigl[ G'(m)+2m^2 G''(m)\bigr]\Gamma+2m^2G'(m)
\bigl[ {\displaystyle b\over \displaystyle 2} G'(m)-1  \bigr]
\end{array}
\right), 
\eea
where
\be
\label{III2}
\Gamma=a+{b \over 2}\varphi^2 + \frac{b}{2}G(m)-m^2.
\ee
Now the stability analysis reduce to a discussion about the signs of
the eigenvalues of $A$ at $u(\bar{\varphi},\bar{m})$, being
$\bar{\varphi}$ and $\bar{m}$ the solutions of
(\ref{II4})-(\ref{II5}).   

\indent Section II has showed  that when $\Gamma=0$, the system
might have two solutions at $\varphi=0$ and $b\bar\varphi^2=3\bar
m^2$. For the former case $A$ is diagonal,
\bea
\label{III3}
 A &=&
\left(
\begin{array}{cc}
\bar m^2 & 0 \\
 0       &
2\bar m^2G'(\bar m)
\bigl[ {\displaystyle b\over \displaystyle 2} G'(\bar m)-1  \bigr]
\end{array}
\right).
\eea
Hence $\varphi$ and $m$ are the eigenvectors of $A$. In addition, 
$\bar m^2$ is the oscillation mode in the $\varphi$ direction. 
Recalling now the renormalization
condition (\ref{II12})-(\ref{II13})   
with $\alpha=1$ we get the 
following expression for the eigenvalue in the $m$ direction,
\be
\label{III4}
2\bar m \Bigl(-\frac{1}{g_{\mu}}
+\frac{1}{16\pi^2}\log\frac{\bar m^2}{\mu^2}\Bigr) \, .
\ee
The system is stable if
\be
\label{III5}
\frac{1}{g_{\mu}} < \frac{1}{16\pi^2}\log\frac{\bar m^2}{\mu^2} \, .
\ee
This is the result we have obtained in I from the RPA analysis.

\indent
For the broken phase the stability matrix is no longer diagonal,
\bea
\label{III6}
A&=&
\left(
\begin{array}{cc}
\bar m^2 & b\varphi \bar{m} G'(\bar{m}) \\
b\varphi \bar{m} G'(\bar{m})       &
2\bar{m}^2G'(\bar{m})
\bigl[ {\displaystyle b\over \displaystyle 2} G'(\bar{m})-1  \bigr]
\end{array}
\right).
\eea
Its determinant is equal to
\be
\label{III7}
{\rm det}A
=-2\bar m^4 G'(\bar{m})\Bigl( bG'(\bar m)+1 \Bigr) 
= 2\bar m^4\Bigl(-\frac{1}{g_{\mu}}+\frac{1}{16\pi^2}
\log\frac{\bar m^2}{\mu^2}\Bigr),
\ee
where we have used the relation $b\bar\varphi^2=3\bar m$ and 
the renormalization
condition (\ref{II12})-(\ref{II13}) with $\alpha=-2$
in the second equality.  
The criterion of minimum requires the eigenvalues to be positive.  
A straighforward calculation yields
\bea
\lambda^{+}&=&\bar m^2L_{\mu}
            \Bigl\{ 3+{\cal O}\Bigl(\frac{1}{L_{\mu}}\Bigr) \Bigr\}, 
\label{III8} 
\\
\lambda^{-}&=&\frac{2\bar m^2}{3L_{\mu}}
 \Bigl(-\frac{1}{g_{\mu}}+\frac{1}{16\pi^2}\log\frac{\bar m^2}{\mu^2}\Bigr)
 +{\cal O}\Bigl(\frac{1}{L_{\mu}^2}\Bigr).
\label{III9} 
\eea
%
%
The eigenvalue $\lambda^{+}$ is always positive while $\lambda^{-}>0$
if the Eq.(3.5) is satisfied. The result shows that $u(\varphi,m)$ is quite
singular at this critical point, because $\lambda^{-}$ is a zero mode
for the limit of infinite cutoff and  the curvature
correspondent to the eigenvalue $\lambda^{+}$ is infinitely sharp. 
%
%
%
%
\renewcommand{\theequation}{4.\arabic{equation}}
\setcounter{equation}{0}
\begin{center}
{\bf IV. RPA Equations}
\end{center}

\indent
Section II discussed in detail the possible minima of ${\cal H}$
as a function of the variational parameters. In this and next sections
we will investigate near equilibrium dynamics around the stationary points. 
In I we have obtained the RPA equations by linearizing the time-dependent
variational equations. Here we will procede differently: we first expand
the hamiltonian, ${\cal H}$, around the  stationary points and then the
RPA equations are obtained from Hamilton's equation with the new
hamiltonian 
[6].
Of course these two approachs are equivalent,
but the method discussed here has advantage of eliminating the step of
introducing a new auxiliary variable $\delta v$ (see Section IV of I).

\indent
Let us first consider fluctuations around 
$\bar{\varphi}$ and $\bar{m}$, which are generic solutions obtained
in the previous section for the uniform system,
\bea
\label{IV1} 
\delta\phi ({\bf x},t)&\equiv &\phi ({\bf x},t)-\bar{\varphi} \, , \\
\label{IV2} 
\delta G({\bf x,y},t) &\equiv &G({\bf x,y},t) - \bar{G}({\bf x,y})\, . 
\eea
They define the one- and two-particle wavefunctions  respectively.
These quantities and their canonical conjugate momenta, 
$\pi_{0}$ and $\Sigma$, are assumed to be small  in our 
approximation.
Next, we expand the hamiltonian, Eq.(1.8), 
up to second order in $\delta\phi$, $\pi_{0}$, $\delta G$ and $\Sigma$,
\bea
\label{IV3} 
{\cal H}_{RPA}
&=&\int_{\bf x} \left(
\frac{1}{2}\bigl[ a + \frac{b}{2} \bar{G}({\bf x,x}) \bigr]\bar{\varphi}^2 
         +\langle {\bf x}\mid\
       \frac{1}{8}\hat{\bar G}^{-1}
      +\frac{1}{2}\hat{p}^2 \hat{\bar G} \mid {\bf x}\rangle 
          +\frac{1}{2b}\bigl[ a+\frac{b}{2}\bar G({\bf x,x}) \bigr]^{2} 
             +\frac{b}{24}\bar{\varphi}^4  \right) \nonumber \\
&+&\int_{\bf x}\left(
\Bigl[ a +\frac{b}{6}\bar{\varphi}^2+\frac{b}{2}\bar{G}({\bf
  x,x})\Bigr] \bar{\varphi} 
            \right) \delta\phi ({\bf x,x},t) \nonumber \\
&+&\int_{\bf x,y} \left(
 {1\over 2} \Bigl[ -\nabla^2_{\bf x}+a+{b \over 2}\bar{\varphi}^2
+ \frac{b}{2}\bar G({\bf x,x})\Bigr]\delta({\bf x-y}) 
- \frac{1}{4}\bar G^{-2}({\bf x,y}) 
                 \right) \delta G ({\bf x,y},t) \nonumber \\
&+&{1\over 2}\int_{\bf x} \pi_{0}^2 ({\bf x},t)  
 +2\int_{\bf x,y,z}\Sigma({\bf z,x},t)\bar G({\bf x,y})\Sigma({\bf
   y,z},t)
      \nonumber \\
&+&\int_{\bf x}                    
 {1\over 2} \Bigl[ -\nabla^2_{\bf x}+a+{b \over 2}\bar{\varphi}^2
+\frac{b}{2}\bar G({\bf x,x})\Bigr]\delta\phi^2({\bf x},t) 
+{b\bar{\varphi}\over 2}\int_{\bf x}
\delta\phi ({\bf x},t)\delta G({\bf x,x},t)    \nonumber \\
&+& 
  {b\over 8}\int_{\bf x} \delta G({\bf x,x},t)\delta G({\bf x,x},t)  
+{1\over 8}\int_{\bf x}
\langle {\bf x}\mid 
\hat{\bar G}^{-1}\delta\hat G\hat{\bar G}^{-1}\delta\hat G\hat{\bar G}^{-1} 
               \mid {\bf x} \rangle \, .
\eea
The first term of this expression is the time-independent part of the
hamiltonian. The second and  third terms vanish because of 
the equilibrium conditions, (2.4) and (2.5). The Eq. (2.6) can also be
used to simplify the sixth term. 
The matrix element of the last term can be written explicitly in 
momentum space as
\bea
\label{IV4} 
&&{1\over 8}\int_{\bf x} \langle {\bf x}\mid 
\hat{\bar G}^{-1}\delta{\hat G}\hat{\bar G}^{-1}\delta\hat G\hat{\bar G}^{-1} 
                                      \mid {\bf x} \rangle \nonumber \\
&&={1\over 2}\int_{\bf p_1,p_2}
 \omega_{\bf p_1}\omega_{\bf p_2} (\omega_{\bf p_1}+\omega_{\bf p_2})
 \delta G ({\bf p_1,p_2},t)\delta G (-{\bf p_1},-{\bf p_2},t)
\eea
where $\omega_{\bf p}$ is a matrix element of the operator in momentum
space,
$$
        \langle {\bf p}_1\mid\hat\omega\mid -{\bf p}_2\rangle
        \equiv \langle {\bf p}_1\mid\
        \sqrt{\hat p^2+\bar m^2}\mid -{\bf p}_2\rangle
        =\delta ({\bf p}_1+{\bf p}_2)\omega_{\bf p}.
$$
%
It is useful to introduce the relative and total momentum coordinates
given respectively by
\bea
\label{IV5} 
{\bf p} &\equiv & ({\bf p}_1-{\bf p}_2)/2 \, ,\\
\label{IV6} 
{\bf P} &\equiv & ({\bf p}_1+{\bf p}_2)\, .
\eea
Thus, from these remarks the Eq.(4.3) can be rewritten, in
terms of  ${\bf p}$, ${\bf P}$ coordinates, as  
\bea
{\cal H}_{RPA}
\label{IV7} 
&=&{\cal H}_{EST}+{1\over 2}\int_{\bf P} \pi_{0}({\bf P},t)\pi_{0}(-{\bf P},t)
  +\int_{\bf p,P}\Sigma({\bf p,P},t)
   \Bigl( {1\over 2\omega_{+}} + {1\over 2\omega_{-}} \Bigr)
                 \Sigma(-{\bf p},-{\bf P},t) \nonumber \\
&+&\int_{\bf P}{1\over 2}                    
 \Bigl[ {\bf P}^2+ \bar m^2 \Bigr]\delta\phi({\bf P},t)\delta\phi(-{\bf P},t) 
\nonumber \\
&+&{b\bar\varphi\over 4}\int_{\bf p,P}
   \Bigl[ \delta\phi ({\bf P},t)\delta G({\bf p},-{\bf P},t)
   +\delta\phi (-{\bf P},t)\delta G({\bf p},{\bf P},t)\Bigr] \nonumber \\
&+&{b\over 8}\int_{\bf p,p',P} \delta G ({\bf p,P},t)
                               \delta G ({\bf p}',-{\bf P},t)
                               \nonumber \\ 
&+&{1\over 2}\int_{\bf p,P}
  \omega_{+}\omega_{-} (\omega_{+}+\omega_{-})\delta G({\bf p,P},t) 
                                      \delta G(-{\bf p},-{\bf P},t)\, ,
\nonumber \\
\eea
%
where $\omega_{\pm}=\sqrt{({\bf p} \pm {\bf P})^2+\bar m^2}$.
It is convenient to make the following changes of variables for the
further use,
\bea
\label{IV8} 
\Sigma ({\bf p},{\bf P},t) h({\bf p},{\bf P},t)
& \equiv &\Pi ({\bf p},{\bf P},t) \, ,\\
\label{IV9} 
\delta G({\bf p},{\bf P},t) h^{-1}({\bf p},{\bf P},t)
& \equiv &\rho ({\bf p},{\bf P},t)\, , 
\eea
being
\be
\label{IV10} 
h({\bf p,P})=\sqrt{\omega_{+}+\omega_{-}
\over 2\omega_{+}\omega_{-}} \,   \, .
\ee
%
%
%
Thus,
\bea
\label{IV11} 
{\cal H}_{RPA}
&=&H_{EST}+{1\over 2}\int_{\bf P} \pi_{0}({\bf P},t)\pi_{0}(-{\bf P},t)
          +{1\over 2}\int_{\bf p,P}\Pi ({\bf p,P},t)
                                   \Pi(-{\bf p},-{\bf P},t) \nonumber \\
&+&\int_{\bf P}{1\over 2}                    
 \Bigl[ {\bf P}^2+\bar m^2 \Bigr]\delta\phi({\bf P},t)\delta\phi(-{\bf P},t) 
        \nonumber \\
&+&{b\bar\varphi\over 4}\int_{\bf p,P} h({\bf p},{\bf P})  
\Bigl[ \delta\phi({\bf P},t)\delta\rho({\bf p},-{\bf P},t)
+\delta\phi (-{\bf P},t)
                  \delta\rho({\bf p},{\bf P},t)\Bigr]\nonumber \\
&+&{b\over 8}\int_{\bf p,p',P} 
 \Bigl( h({\bf p,P}) \delta\rho ({\bf p,P},t) \Bigr)
 \Bigl( h({\bf p}',{\bf P}) \delta\rho ({\bf p}',-{\bf P},t)\Bigr)\nonumber \\
&+&{1\over 2}\int_{\bf p,P}
   \bigl( \omega_{+}+\omega_{-}\bigr)^2
   \delta\rho ({\bf p,P},t)\delta\rho (-{\bf p},-{\bf P},t) \, .
\eea
From this one can get the linearized equations of motion by following
directly Hamiltion's equations [1]:
\bea
\label{IV12} 
\delta\dot\phi ({\bf P},t) 
&=&\pi_{0} (-{\bf P},t) \, , \\
\label{IV13} 
\dot\pi_{0} ({\bf P},t)
&=&
-\Bigl[ {\bf P}^2+\bar m^2 \Bigr]\delta\phi(-{\bf P},t) 
-{b\bar\varphi\over 2}
 \int_{\bf p}h({\bf p},{\bf P})\delta\rho ({\bf p},-{\bf P},t) \, , \\
\label{IV14} 
\delta\dot\rho ({\bf p},{\bf P},t)
&=&\Pi (-{\bf p},-{\bf P},t) \, ,\\
\label{IV15} 
\dot\Pi ({\bf p},{\bf P},t)  
&=&
- (\omega_{+} +\omega_{-})^2\delta\rho (-{\bf p},-{\bf P},t)
-{b\over 4} h({\bf p,P})
   \int_{\bf p} h({\bf p',P}) \delta \rho ({\bf p}',-{\bf P},t) \nonumber\\
&-&{b\bar\varphi\over 2} 
\int_{\bf p} h({\bf p,P}) \delta\phi (-{\bf P},t) \, .
\eea
%
Eliminating the canonical momenta $\pi_{0}$ and $\Pi$ we arrive
finally at
\be
\label{IV16} 
\delta\ddot\phi ({\bf P},t) + \omega^2_{\bf P}\delta\phi({\bf P},t) 
  +{b\bar\varphi\over 2} \int_{\bf p'} h({\bf p',P})\delta\rho ({\bf p',P},t) 
  =0\, ,
\ee
\be
\label{IV17} 
\delta\ddot\rho+(\omega_{+} +\omega_{-})^2\delta\rho ({\bf p},{\bf P},t)
               +{b\over 4} h({\bf p,P})
   \int_{\bf p} h({\bf p',P}) \delta \rho ({\bf p}',{\bf P},t)
   \nonumber \\
   +{b\bar\varphi\over 2} h({\bf p,P}) \delta\phi ({\bf P},t)
   =0 .
\ee
These are linear oscillator equations as usual in RPA treatment. The
solutions for this problem involves determining the modes of
oscillation and their eigenfrequences.

\indent We have seen in Sec.II that $\bar\varphi$ may have solutions at
$\varphi=0$ and $\varphi^2=3\bar m^2/b$. Note that in the symmetric
phase the two equations decouple. The two-particle equation
reduces to Eq.(4.20) of I while the one-particle equation becomes 
simple oscillator equations, one for each $\bf{P}$ with frequence 
$\omega_{\bf P}$,
\be
\label{IV18} 
\delta\ddot\phi ({\bf P},t) + \omega^2_{\bf P} \delta \phi ({\bf
  P},t)=0\, .
\ee
For the solution $\bar\varphi^2=3\bar m^2/b$, however, 
the coupling between the 
one-particle and two-particle wavefunction is nontrivial, we shall discuss the
solution of this problem in the next section.

\renewcommand{\theequation}{5.\arabic{equation}}
\setcounter{equation}{0}
\begin{center}
\indent {\bf V. Small Oscillation Equation and Scattering Problem}
\end{center}

\indent
Let us first remove the time dependence of 
(\ref{IV16})-(\ref{IV17})  by writting
\bea
\label{V1} 
\delta\phi({\bf P},t)&=&\delta\phi^{(0)}({\bf P})\cos[\Omega({\bf
  P})t] \, ,\\
\label{V2} 
\delta\rho({\bf p,P},t)&=&\delta\rho^{(0)}({\bf p,P})\cos[\Omega({\bf
  P})t]\, ,
\eea
\noindent
where $\delta\phi^{(0)}$ and $\delta\rho^{(0)}$ are the amplitude of
oscillation for the one- and two-particle modes 
respectively and $\Omega({\bf P})$ 
are the eigenfrequences (see section IV-b of I for interpretation of the
wavefunctions). Substituting these into (\ref{IV16})-(\ref{IV17}) we have
the following eigenvalue problem:
\be
\label{V3} 
  \omega^2_{\bf P}\delta\phi^{(0)}({\bf P})+ 
 {b\bar\varphi\over 2}\int_{\bf p'}h({\bf p',P})\delta\rho^{(0)}({\bf p',P}) 
  =\Omega^2\delta\phi^{(0)}({\bf P})\, ,
\ee
\bea
\label{V4} 
  (\omega_{+} +\omega_{-})^2\delta\rho^{(0)}({\bf p},{\bf P})
  &+&{b\over 4} h({\bf p,P})
   \int_{\bf p} h({\bf p',P}) \delta \rho^{(0)} ({\bf p}',{\bf P})\nonumber\\
  &+&{b\bar\varphi\over 2} h({\bf p,P}) \delta\phi^{(0)}({\bf P})
  =\Omega^2\delta\rho^{(0)} ({\bf p},{\bf P}).
\eea
Since the amplitudes of oscillations are wavefunctions of quantum particles  
it is interesting to treat this system
as a coupled channel scattering problem with appropriate boundary
conditions. Henceforth, we shall use $\alpha$ and $\beta$ to denote 
the one and two-particle channel respectively. In the 
following discussion we will consider the cases separately.

\smallskip
\indent { a. $\alpha \longrightarrow \alpha$ }
\smallskip

\indent In this process one has a incident wave of $\delta\phi^{(0)}$;
it couples  to $\delta\rho^{(0)}$ through the term 
$b\bar{\varphi}\int h\delta\rho^{(0)}$ and reemits $\delta\phi^{(0)}$
at the exit channel. Thus, 
we can formally solve (\ref{V4}) as follows:
\be
\label{V5} 
\delta\rho^{(0)}({\bf p,P})=
{1\over \Omega^2-(\omega_{+}+\omega_{-})^2+i\epsilon}
\Bigl[{b\over 4} h({\bf p,P})
   \int_{\bf p} h({\bf p',P}) \delta \rho ({\bf p}',{\bf P},t)
  +{b\bar\varphi\over 2} h({\bf p,P}) \delta\phi ({\bf P},t)
\Bigr] \, .
\ee
This expression includes the boundary condition that there is no
indent wave of $\delta\rho^{(0)}$. The term  
$\int_{\bf p}h({\bf p,P})\delta\rho^{(0)} ({\bf p,P})$ couples
$\delta\rho^{(0)}$ of different relative momenta. We can solve for this term
by multiplying (\ref{V5}) by $h({\bf p,P})$ and 
integrates it with respect to ${\bf p}$,
\be
\label{V6} 
\int_{\bf p'} h({\bf p',P})\delta\rho^{(0)}({\bf p,P})
={2\bar\varphi I_{\bf P}(\Omega)
\over\displaystyle  {2\over b}
-I_{\bf P}(\Omega)}\delta\phi^{(0)}({\bf P})\, ,
\ee
where
\be
\label{V7} 
I_{\bf P}(\Omega)=\int_{\bf p}{h^2({\bf p,P})
                  \over\Omega^2-(\omega_{+}+\omega_{-})^2+i\epsilon}\,
                  .
\ee
The details of the computation of this integral can be found 
in the appendix B of I. For the present analysis it is
sufficient to say
\be
\label{V8} 
I_{\bf P}(\Omega)=-L_{\mu}+F_{\bf P}(\Omega)\, ,
\ee
where $L_{\mu}$ is the logarithmic divergent term defined in
(\ref{II12}) and $F_{\bf P}(\Omega)$ the finite part of 
$I_{\bf P}(\Omega)$ [See (6.3)-(6.6) for $I_{\bf P}(\Omega)$].  
Substituing now (\ref{V6}) into (\ref{V3}) we have
\be
\label{V9} 
\Bigl[\Omega^2-\omega^{2}_{\bf P}
-{b\bar\varphi^{2}I_{\bf P}(\Omega)\over 
  2/b-I_{\bf P}(\Omega) }\Bigr]\delta\phi^{(0)}({\bf P})=0\, .
\ee
For the asymmetric solution, $b\bar\varphi^2=3\bar{m}^2$, and 
(\ref{V9}) becomes
\be
\label{V10} 
\Bigl[\Omega^2-{\bf P}^2
-\bar{m}^2\Bigl(1+{3I_{\bf P}(\Omega)\over 
  2/b-I_{\bf P}(\Omega) }\Bigr)\Bigr]\delta\phi^{(0)}({\bf P})=0\, .
\ee
Using the renormalization condition (\ref{II13})
and (\ref{V8}) one arrives at
\be
\label{V11} 
\Bigl[\Omega^2-{\bf P}^2
-\bar{m}^2{\cal O}\Bigl({1\over L_{\mu}}\Bigr)\Bigr]
\delta\phi^{(0)}({\bf P})=0\, .
\ee
Notice that the effect of the coupling $\delta\phi^{(0)}$ to the 
two-particle modes switch
the mass of the particle. In the limit of infinity cutoff its
effective mass goes to zero.

\smallskip
\indent { b. $\beta \longrightarrow \beta$}
\smallskip

\indent It is also an elastic scattering process,
where the entrance channel as
well as the exit channel exhibit two-particle wave
$\delta\rho^{(0)}$. Analogously to the previous discussion we first
solve (5.3) as
\be
\label{V12} 
\delta\phi^{(0)}({\bf P})=
{b\bar\varphi/2\over \Omega^2-\omega_{\bf P}^2+i\epsilon}
\int_{\bf p'}h({\bf p',P})\delta\rho({\bf p',P})\, .
\ee
%
Using this and (5.4) we get the
following integral equation for $\delta\rho^{(0)}$:
%
%
\be
\label{V13} 
\delta\rho^{(0)} ({\bf k},{\bf p,P};\Omega)
=\gamma_{\beta}\delta ({\bf k}-{\bf p}) 
+{b\over 4}
 \Bigl({\displaystyle 1+{b\bar\varphi^{2}
       \over \Omega^2-\omega_{\bf P}^{2} +i\epsilon }
  \over \Omega^2-(\omega_{+} + \omega_{-})^2+i\epsilon} \Bigr)
        h({\bf p,P})\int_{\bf p'}h({\bf p',P})
        \delta\rho^{(0)} ({\bf k},{\bf p',P};\Omega) \, .
\ee
In this expression
$\bf k$ is the relative momentum of the two incident meson;
$\gamma_{\beta}$ indicates a phase factor to keep $\delta\rho^{(0)}$
real and the subindex $\beta$ denotes the source of incident wave.
From (\ref{V13}) one finds immediately 
$$
\label{V14} 
\int_{\bf p}h({\bf p,P})\delta\rho^{(0)} ({\bf p,P})
=\gamma_{\beta}{b\over2}{1\over\Delta^{+}_{\bf P}(\Omega)}
 h({\bf p,P})\, ,
$$
where
\be
\Delta^{+}_{\bf P}(\Omega)
={2\over b}-\Bigl( 1+{b\bar{\varphi}^{2}
\over \Omega^2-\omega_{\bf P}^{2}+i\epsilon}\Bigr) I_{\bf P}(\Omega)\, .
\ee
Using this result in (\ref{V13}) we find
\be
\label{V15} 
\delta\rho^{(0)} ({\bf k},{\bf p,P};\Omega)
=\gamma_{\beta}\delta ({\bf k}-{\bf p})
+{1\over \Omega^2-(\omega_{+} + \omega_{-})^2+i\epsilon}
\gamma_{\beta}h({\bf k,P})
\Xi ({\bf P},\Omega)
h({\bf p,P}) 
\ee
with
\be
\label{V16} 
\Xi ({\bf P},\Omega)
=\Bigl(
{\displaystyle 1+
{b\bar\varphi^{2}\over \displaystyle\Omega^2-\omega_{\bf P}^{2}
+i\epsilon} }\Bigr)
{1 \over
\displaystyle \Delta^{+}_{\bf P}(\Omega) } \, .
\ee
Of course, this result 
reduce to Eq.(5.5) of I in the case of $\bar\varphi=0$. For the broken
solution, we can rewrite (\ref{V16})
in terms of renormalized parameters as
\bea
\label{V17} 
\Xi ({\bf P},\Omega)
={\displaystyle \Omega^2-{\bf P}^2+2\bar{m}^2
\over
  \displaystyle
3L_{\mu}(\Omega^2-{\bf P}^2)
-{2\over g_{\mu}}(\Omega^2-{\bf P}^2-\bar{m}^2)
-F({\bf P},\Omega)[\Omega^2-{\bf P}^2+2\bar{m}^2]
}\, .
\eea
Notice that $\Xi\rightarrow L_{\mu}^{-1}$ for large $\Lambda_{\bf p}$, 
which means that the
scattering waves is infinitesimaly small. Therefore, 
{\bf effectively} there is
no interaction between the particles coming from this broken vacuum. 

\smallskip
\indent { c. $\alpha \longrightarrow \beta$}
\smallskip

\indent In this case we want to obtain $\delta\rho^{(0)}$ with
a source of one particle, which is an inelastic processes. 
To do so, (\ref{V3}) is solved as
\be
\label{V18} 
\delta\phi^{(0)}({\bf P,K},, \Omega)=\gamma_{\alpha}\delta({\bf K}-{\bf P})
+{1\over\Omega^2-\omega_{\bf P}^2+i\epsilon}{b\bar\varphi\over2}
\int_{\bf p'}h({\bf p',P})\delta\rho^{(0)}({\bf p',P},\Omega).
\ee
Here we have included an incident wave of $\delta\phi^{(0)}$ with 
moment $\bf K$. Using this
solution in (\ref{V4}) one gets an integral equation 
for $\delta\rho^{(0)}$ with a source in $\alpha$, 
\bea
\label{V19} 
\bigl[-\Omega^2+(\omega_{+}+\omega_{-})^2\bigr]\delta\rho^{(0)}({\bf p,P})
&=&
\gamma_{\alpha}\delta({\bf K}-{\bf P}){b\bar\varphi\over2}h({\bf p',P})
\nonumber \\
+{b\over 4}\Bigl(1
&+&{b\bar\varphi^2\over \Omega^2-\omega_{\bf P}^2+i\epsilon}\Bigr)
 h({\bf p,P})\int_{\bf p'} h({\bf p',P})\delta\rho^{(0)}({\bf p',P}).
\eea
This equation can be solved as usual given
\bea
\label{V20} 
\delta\rho^{(0)} ({\bf K},{\bf p,P};\Omega)
&=&
\gamma_{\alpha}\delta({\bf K}-{\bf P}){b\bar\varphi\over2}h({\bf p',P})
\Bigl[1+\Xi ({\bf P},\Omega)\Bigr] \nonumber \\
&=&
\gamma_{\alpha}\delta({\bf K}-{\bf P}){b\bar\varphi\over2}h({\bf p',P})
\Bigl[ 1+{\cal O}\Bigl( {1\over L_{\mu}} \Bigr)
\Bigl]\, .
\eea
In the second line we have used (\ref{V17}).
Recalling the discussion in the 
section II, we have learned that in the broken solution, the mean-field
$\varphi$ requires a scaling factor. As consequence, we have
essencially $b\bar\varphi\propto  1/\sqrt{L_{\mu}}$, which vanishes
for large cutoff.
Therefore, 
in the continuum limit $\delta\rho$ cannot be observed in this reaction. 

\smallskip
\indent { d. $\beta \longrightarrow \alpha$}
\smallskip

\indent For the last case, the Equation (\ref{V4}) 
is solved with the source in $\beta$,  
\bea
\label{V21} 
\delta\rho^{(0)} ({\bf k},{\bf p,P};\Omega)
&=&\gamma_{{}_\beta}\delta ({\bf k}-{\bf p}) 
+{1\over \Omega^2-(\omega_{+} + \omega_{-})^2+i\epsilon}\times\nonumber\\ 
&& \Bigl[ {b\over 4} h({\bf p,P})\int_{\bf p'}
 h({\bf p',P})\delta\rho^{(0)} ({\bf k},{\bf p',P};\Omega)
 +{b\bar\varphi\over2}h({\bf p,P})
 \Bigr]\, .
\eea
From this, one gets immediately
\be
\label{V22} 
\int_{\bf p} h({\bf p,P})\delta\rho^{(0)}({\bf p,P})
=
{1\over 1-bI_{\bf P}(\Omega)/2}
\Bigl[\gamma_{\beta}h({\bf k})
+b\bar\varphi I_{\bf P}(\Omega)\delta\phi\rho^{(0)}({\bf p})\Bigr]\, .
\ee
Combining this with (\ref{V3}) results
\be
\label{V23} 
\Bigl[\Omega^2-\omega^{2}_{\bf P}
-{b\bar\varphi^{2}I_{\bf P}(\Omega)\over 
  2/b-I_{\bf P}(\Omega) }\Bigr]\delta\phi^{(0)}({\bf P})
 =\gamma_{{}_\beta}b\bar\varphi h({\bf k}){1\over 
  1-bI_{\bf P}(\Omega)/2 }\, .
\ee
Above equation differs (\ref{V9}) from the 
boundary condition for this scattering process. Here
we have incoming waves of $\delta\rho$ characterized by 
$\gamma_{{}_\beta}$ in its rhs. It is proportional to $b\bar\varphi$, or
$\delta\phi^{(0)}\propto 1/\sqrt{L_{\mu}}$. 
Therefore, we conclude that this physical process actually cannot 
happen when the cutoff is taken to infinite. 

\indent In summary, this section has discussed the solution for the RPA
equations in the
context of $b\rightarrow 0^{+}$. The excitations
are interpreted as quantum particles and the RPA equations are
treated as a coupled channel scattering problem. The system of
equations allows analytical solutions and we have obtained the
amplitudes of oscillation which are wavefunction of one and
two-particles. In particular, we have shown that at one-particle channel
the effect of coupling is to reduce the effective mass of the particles
and two-particle channel leads to infinitesimal scattering waves
for large cutoff. Furthermore, the inelastic processes, cases c. and d.,
cannot occur when the renormalized field theory is considered.

%
\renewcommand{\theequation}{6.\arabic{equation}}
\setcounter{equation}{0}
\begin{center}
{\bf VI. Discussion}
\end{center}

\indent From the previous sections and I we have seen that the Gaussian
variational approximation indicates two distinct nontrivial version of
$(\phi^4)_{3+1}$ theories. They are characterized by the bare coupling
to be infinitesimal of the form $b\rightarrow \pm
L_{\mu}^{-1}$ in the continuum limit.  
Although classical intuition and perturbation calculation indicate that
the theory with $b\rightarrow0^{-}$ is unstable,
but preceding analysis and I reveal that it has a well defined symmetric
phase. In particular the renormalized version of the theory allows a
stable vacuum and finite scattering amplitude[1][4]. The other one,
$b\rightarrow0^{+}$, shows a renormalized broken phase solution,
$b\bar{\varphi}^2=3\bar{m}^2$, which is degenerate with the symmetric phase.  
In section III we have discussed the conditions of stability for these
vacuums by studying the eigenvalues of the stability matrix. 
An alternative approach to investigate this
problem is through analyses of the positions of the RPA modes. The method
is more general because it includes inhomogeneous degree of freedom.
In the following discussion we will
consider the cases of symmetric and broken solution  separately.

\smallskip
 \begin{center}
{\bf a. Symmetrical Phase}
\end{center}
\smallskip

\indent {\bf a-1. Massive Theory}

In this case $\delta\phi$ decouples from $\delta\rho$ and (\ref{V3}) becomes
\be
\label{VI1}
(s-\bar{m}^2)\delta\phi^{(0)}=0\, ,
\ee
where we have introduced the covariant variable 
$s\equiv \Omega^2-{\bf P}^2$ [7].
Thus, this channel yields a simple free-particle spectrum. 
The Eq.(5.4), on the other hand, describes the nontrivial sector of
this vacuum where one has excitations of two 
particles. 

\indent This case was explored in I 
and here we will summarize the main results. 
In order to get the spectrum we write the scattering matrix for this
process as
\be
\label{VI2}
        {\bf T}({\bf p,p',P;} \Omega)
        =h({\bf p},{\bf P}) 
{1\over\displaystyle {2\over b}-I_{\bf P}(\Omega)}
         h({\bf p'},{\bf P})\, .
\ee
In appendix B of I we have performed the calculation of 
$I_{\bf}(\Omega)$. In terms of covariant variable $s$ the result is 
\be
\label{VI3}
I(s)=-L_{ m}-{1\over 16\pi^2}f(s)
     -\theta(s-4\bar{m}^2){i\over 16\pi}\sqrt{{s-4\bar m^2\over s}}\, ,
\ee
where, $L_{m}=1/8\pi^2\log(2\Lambda_{\bf p}/e\bar{m})$ and
\bea
\label{VI4}
        f(s)&=&2+\sqrt{{s-4\bar m^2\over s}}
                \log {\sqrt{s}-\sqrt{s-4\bar m^2}
                      \over
                      \sqrt{s}+\sqrt{s-4\bar m^2}}
        \qquad s>4\bar m^2,  \\
\label{VI5}
        f(s)&=&2-2\sqrt{{-4\bar m^2\over s}}
                \tan^{-1}\sqrt{{ s\over -4\bar m^2}}
        \qquad 0< s <4\bar m^2,  \\
\label{VI6}
        f(s)&=&2+\sqrt{{s-4\bar m^2\over s}}
                \log {\sqrt{s-4\bar m^2}-\sqrt{s}
                      \over
                      \sqrt{s-4\bar m^2}+\sqrt{s}}
        \qquad s < 0    
\eea
(cf. fig.3 of I). Depending on the value of $s$ the system can have
different dynamical behavior. 
When $s<0$ the system is unstable and for $s>4\bar{m}^2$ it
has a continuum spectrum. In the interval of $0<s<4\bar{m}^2$ the
system may present stable bound states if one finds $s_{_{B}}$ such as
\be
\label{VI7}
\Delta^{+}(s_{_{B}})={2\over b}+L_{m}
+{f(s_{_B})\over16\pi^2}=0\, .
\ee
In this interval, $0\le f(s)\le 2$, this  
equation has solution if, only if $b\rightarrow -L_{m}^{-1}$. In
particular one can choose the renormalization condition used for the
static equations, i.e., (\ref{II12})-(\ref{II13}).
In this way, we can show that this theory will
result a stable bound state when the renormalized coupling 
constant $g_{m}<-1/8\pi^2$ (cf. section V of I).

\indent {\bf a-2. Massless Theory}

\indent  In section II we have shown that with the renormalization
condition $a+b\Lambda/16\pi^2=0$ the system has a massless
solution. In particular the broken phase is degenerate with the
symmetric phase. In this case it is still convenient
to use $\bar m$ as the mass scale defined by the relation
(\ref{II21}). Using these ingredients we get
\be
\label{VI8}
\Delta^{+}(s<0)
={2\over b} - I(s) \nonumber\\
=-{\alpha\over 8\pi^2}\log{2\Lambda_{\bf p}\over\sqrt{e}\bar{m}}
 +{1\over 16\pi^2}\log{\Lambda_{s}\over |s|}\, .
\ee
In this expression $\alpha$ labels the two theories as we have
discussed before.\\ 
\indent Case 1: $\alpha=1$ ($b\rightarrow0^{-}$)\\ 
\indent The equation (\ref{VI8}) becomes
\be
\label{VI8a}
\Delta^{+}(s<0)
=-{1\over 8\pi^2}\log{2\Lambda_{\bf p}\over e\bar{m}}
 +{1\over 16\pi^2}\log{\Lambda_{s}\over |s|}
= {1\over 16\pi^2}\log{e^2\bar{m}^2\over |s|}\, ,
\ee
where we have used the relation $\Lambda_{s}=(2\Lambda_{\bf
p})^2$. The system is stable if $\Delta^{+}(s<0)\neq 0$ for
any $s$. However, from (\ref{VI8}) one can always find $s_{i}$ such as
$\Delta^{+}(s_{i})=0$. Therefore the vacuum is unstable when
$b\rightarrow0^{-}$. 

\indent Case 2: $\alpha=-2$ ($b\rightarrow0^{+}$)  

\indent In this case the broken phase collapses to the symmetric
vacuum. The stability analysis is analogous to the previous one:
 \be
\label{VI8b}
\Delta^{+}(s<0)
= {2\over 8\pi^2}\log{2\Lambda_{\bf p}\over e\bar{m}}
 +{1\over 16\pi^2}\log{\Lambda_{s}\over |s|}
= {1\over 16\pi^2}\log{\Lambda_{s}^3\over e^4\bar{m}^4|s|}\, .
\ee
The solution for $\Delta^{+}(s)=0$ is
\be
|s|={\Lambda^3_{s}\over e^4\bar m^4}\, .
\ee
This solution is unphysical and has to be eliminated. Therefore, the 
system is stable in this case.

\indent Thus, the Gaussian variational approximation indicates that
$(\phi^4)_{3+1}$ with $b\rightarrow 0^{-}$ might be
a nontrivial theory. It has a well defined potential for the sector of
$(\varphi=0,m)$ of the variational space. For certain range of
renormalized coupling constant the theory allows stable vacuum. Its
excitations yield one (free) and two-particle modes. The two-meson
equation leads to a single bound state and the scattering amplitude in
the continuum. We also find a massless solution with the
renormalization condition $a+b\Lambda^2/16\pi^2=0$. 
The RPA analyses indicate that the solution is: i)unstable 
if $b\rightarrow0^{-}$, which
suggest that the true vacuum is not homogeneous; ii)stable for 
theory with $b\rightarrow0^{+}$. 

\smallskip
{\bf b. Broken Phase}
\smallskip

\indent In this case we have the one- and two-particle wavefunction 
coupled through of $b\bar\varphi$. In order to discuss the  RPA
equations  (\ref{V10}) is rewritten 
in terms of covariant variable $s$ as
\be
\label{VI9}
\Bigl[s-\bar{m}^2\Bigl(1+{3I^{+}(s)\over 
2/b-I^{+}(s) }\Bigr)\Bigr]\delta\phi^{(0)}=0\, .
\ee
In the interval of
$0<s\leq4\bar{m}^2$, $I^{+}(s)$ is real and the system may present a
stable state if one finds a solution $s_{_{B}}>0$ such as
\bea
\label{VI10}
s_{_{B}}&=&\bar{m}^2\Bigl[1+{3I^{+}(s_{_{B}})
\over\displaystyle 2/b-I^{+}(s_{_{B}})}\Bigr] \nonumber \\ 
        &=&\bar{m}^2\Bigl[1-{3
\over\displaystyle 1+{2\over b(L_{m}+f(s_{_{B}})/16\pi^2)}} \Bigr]
\, ,
\eea
where we have used (\ref{VI3}). 
Using now (2.12) with $\alpha=-2$,  we arrive at the follwing equations for
$s_{_{B}}$:
\be
\label{VI11}
s_{_{B}}={\bar m^2\over24\pi^2L_{\mu}}\Bigl[-{1\over\bar g_{\mu}}
        +\log{\bar{m}^2\over\mu^2}+f(s_{_{B}})\Bigr]\, .
\ee
Thus, one may have a solution for $s_{_{B}} \rightarrow
L_{\mu}^{-1}$ and it is stable if the bracket is positive. From fig.3 of
I one can see that  $f(s)$ goes to zero for small $s$. Therefore, we
might neglect the last term as first approximation to yield
\be
\label{VI12}
s_{_{B}}\approx{\bar m^2\over24\pi^2L_{\mu}}\Bigl[-{1\over\bar g_{\mu}}
+\log{\bar{m}^2\over\mu^2}\Bigr]\, .
\ee
This is precisely the eigenvalue found in Sec. III (cf.Eq.(3.9)).

\indent When $s>4\bar{m}^2$ $I^{+}(s)$ has an imaginary part and 
(\ref{VI12}) becomes, after the renormalization,
%
\be
\label{VI13}
\Bigl[
s-{\bar m^2\over24\pi^2L_{\mu}}\Bigl(-{1\over\bar g_{\mu}}
+\log{\bar{m}^2\over\mu^2}+f(s)
+i\pi\sqrt{s-4\bar m^2\over s}\Bigr)\Bigr]\delta\phi^{(0) }=0\, .
\ee
Hence, we have continuum spectrum in this case. 

\indent It is still illustrative to see the above results from the elastic
channel, i.e., $\beta\rightarrow\beta$, where physics is described by the
integral equation (\ref{V19}). Note that the 
potential term is also separable and
the scattering $\bf T$ matrix can be obtained as usual [8]:
\be
\label{VI14}
        {\bf T}({\bf p,p',P;} \Omega)
        =h({\bf p},{\bf P}) 
         \Xi ({\bf P;} \Omega)
         h({\bf p'},{\bf P})\, ,
\ee
where $\Xi$ is given by (\ref{V22}). In terms of $s$ it can be
rewritten as
\be
\label{VI15}
\Xi(s)=\Bigl(1+{3\bar{m}^2\over
    s-\bar{m}^2}\Bigr){1\over\Delta^{+}(s)}\, ,
\hspace{1.0cm}
\Delta^{+}(s)={2\over b}-\Bigl(1+{3\bar{m}^2\over s
-\bar{m}^2}\Bigr)I^{+}(s)\, .
\ee
From this viewpoint several dynamical behavior is described by
denominator of the scattering matrix $\bf T$, i.e.,
$\Delta^{+}(s)=0$. In particular, the solution found in the equation
(\ref{VI12})  can be seen as a bound state of two massless  
particles. It is stable for $g_{m}<0$. 
Another intereting point is to obsever $\Xi(s)$ when
$s>4\bar{m}^2$:
\be
\Xi(s)=
{s+2\bar{m}^2 \over \displaystyle
3s\bigl(L_{\mu}-{1\over g_{\mu}}\bigr)+
(s+2\bar{m}^2)\Bigl\{{1\over g_{m}}
+{1\over16\pi^2}\Bigl[f(s)
+i\pi\sqrt{s-\bar{m}^2\over\bar{m}^2}\Bigr]\Bigr\}}
\ee
In this range of energy the spectrum is continuum, but
the scattering amplitude vanishes as $1/L_{\mu}$. In other words, the
particles {\bf effectively} do not interact, but in a very non-trivial way.

\indent In summary, we have obtained RPA equations for $\phi^4$ field
theory by considering near equilibrium dynamics  about the critical
points of the gaussian parameter space. A simple analytical solution
can be obtained and it is a nonperturbative method to investigate one
and two-meson physics. Using this framework we have investigate the
stability of the theory and shown that
in the continuum limit two
distinct version of $(\phi^4)_{3+1}$ are viable. One has $b\rightarrow
0^{-}$ and its renormalized theory is stable at the symmetric phase. The
other, $b\rightarrow 0^{+}$, indicates spontaneous symmetric breaking.
However, the excitation fields {\bf effectively} do not interact
leading to infinitesimal scattering amplitude [9]. 
Finally, we comment that the techniques developed 
here are general and can be readily extented  to 
other relativistic
field models [10] and low energy many-body problems
such as Boson condensation [11]. 

\newpage
\noindent\Large{\bf Aknowledgement}
\normalsize
\baselineskip=24pt

\noindent The authors are grateful
to Prof. R. Jackiw for usuful discussions. One of us (C-Y. L.) thanks
the members of the Center for Theoretical Physics at MIT, where part
of this work was performed, for their hospitality and support. He also
benefited from numerous conversations with Prof. A. F. R. de Toledo
Piza concerning the RPA results. 

\medskip


\end{document}